\documentstyle[aps,twocolumn]{revtex}

\begin{document}

\twocolumn[\hsize\textwidth\columnwidth\hsize\csname 
@twocolumnfalse\endcsname

\author{Osame Kinouchi\\
DFM - FFCLRP - USP
Av. dos Bandeirantes 3900, CEP 14040-901, 
Ribeir\~ao Preto, SP, Brazil\\ E-mail: osame@dfm.ffclrp.usp.br}

\title{Persistence solves Fermi Paradox but challenges SETI 
projects}

\maketitle

\begin{abstract}
Persistence phenomena in colonization processes
could explain the negative results of SETI search
preserving the possibility of a galactic civilization.
However, persistence phenomena also indicates that search of 
technological civilizations in stars in the neighbourhood 
of Sun is a misdirected SETI strategy.
This last conclusion is also suggested by a weaker form of the Fermi paradox.
A simple model of a branching colonization which includes emergence,
decay and branching of civilizations is proposed. The model could also be used in the
context of ant nests diffusion.
\end{abstract}
\bigskip

]

	Fermi paradox has important scientific and science policy consequences. For example, 
how much money is it reasonable to spend on SETI-like projects? Simply stated, Fermi paradox 
arises from a back envelop calculation about how much time a technological civilization, able to 
perform inter-stellar colonization, needs to diffuse through the entire galaxy. With reasonable 
assumptions, this calculation gives a colonization time of the order of a hundred million years 
\cite{Crawford} \cite{Webb}. 
But since, notwithstanding X-files fans and Erick von Danik\"en, they are clearly not here, one must 
conclude that there is no such galactic civilization. 
	
	I live in Brazil, and sometimes I wonder about such exotic places like New York or Paris 
and the curious costumes of their inhabitants: they do not make gestures when speaking and mix 
sweet food with salty food! Some people drink warm beer and even do not like soccer! And 
their women use very large bikinis! Very irrational behavior, indeed! However, I must recognize 
that there are also a lot of exotic places (and even populations) inside Brazil, which are almost 
never visited or contacted by the global civilization. Suppose that you are a member of a lost 
Amazonian tribe that has never been contacted. Now, it is obvious that a technological civilization 
able to perform travel by air at 1000 km/hour certainly had  time to colonize the entire globe. But 
since they have not reached you (remember, you are a member of a undiscovered Ianomani tribe), 
should you conclude that there is no such global civilization?
	
	A nice view of the global civilization is given by the nocturnal Earth observed from space 
(see figure at \cite{figure}). It is clear from this view not only 
the perverse distribution of global wealth but also the fact that the distribution of technological 
human colonies (commonly called ``cities") is highly non-uniform. Huge areas are not inhabited, 
and even never visited (or visited sporadically only by fanatic explorers). But, despite the 
provincial worldview of lost tribes, the global civilization is there.
	
	The colonization process clearly is not a simple and uniform diffusion process as assumed 
by Fermi calculation. However, even some simple diffusive processes have important properties 
not considered by Fermi paradox advocates. Here I call attention to the existence, in various 
systems, of the so called phenomenon of "persistence", that is, the number of unvisited sites may 
decay not exponentially with time but as a power law. In other words, the probability that a site 
has not been visited (or colonized) by diffusive walkers is given, in certain scenarios, by
\begin{equation}
	P(t) = P_\infty + C t^{-\theta} \:,
\end{equation}

where $P_\infty$ is an asymptotic probability, $C$ is a constant, 
$t$ corresponds to time and $\theta$ is the 
persistence exponent \cite{DB}. 

This formula gives origin to two important facts to be considered by exobiologists. First, 
the probability of a site never being visited, even in the infinite time limit, is a non-zero value P¥. 
Second, even this asymptotic value is reached by a slow (power law) process, not an exponential 
one. If $\theta$ is small, a very long time must be necessary for a site being visited for the first time.
Of course, for long times, these unvisited sites are not typical (however, due to the power 
law behavior, their number is not insignificant). 

Therefore, we must consider whether we are 
typical or we are like the lost Amazonian tribe. That is, what is the probability that Earth is a 
persistent site unvisited by aliens? But now arises an Anthropic-like (or better, probabilistic) 
argument. We surely pertain to an independent, not colonized civilization. The pre-requisite to this 
fact is that we have not been visited (at least, not very much) by aliens, because (as we all know) 
low technology civilizations normally do not survive to the contact with civilizations of more 
advanced technology. Notice that the argument is not circular. What I am saying is that, if we 
would reside in a typical region of the galaxy (that is, one with high traffic) then we would not be 
here considering the Fermi paradox but our civilization or even our Biosphere should be already 
extinct (at least as an independent development). In other words, the conditional probability $
P(\mbox{Earth is typical} | \mbox{we are independent}) $
is very small under the galactic civilization 
hypothesis, leading to the conclusion that we should not expect that our civilization is typical (again 
under this hypothesis). So, Fermi paradox assumptions are misleading.

	Another consequence of taking in account persistence effects is that persistent sites are 
clusterized (like Amazonian forest, Siberia, Australia, Canada, Indonesia islands etc.). This means 
that, if we reside at a persistent site (and we do, because we have not been colonized!) then our 
neighboring stars probably are also persistent sites, that is, do not have inhabited planets or 
technological civilizations. This follows simply from the fact that, if technological civilizations 
developed near the Earth, then they already had the time to colonize us (Fermi paradox is valid in 
small scale, not in the larger one). Therefore, it is not productive, as some SETI projects do, to 
search for other civilizations in neighboring stars, as is not a good strategy for lost tribes to send or 
try to receive smoke and drum signals from neighboring regions. 

I conclude that it is more profitable to concentrate SETI searches in, for example, trying to 
detect for distant stars radio/laser/neutrino signals and artificial features in stellar clusters or nearby 
galaxies. This is not an easy task: for each apparent artificial signal one could think a plethora of 
natural hypothesis to explain it. Think about what should a lost tribe to do for detecting the global 
civilization. The only artificial signals that lost tribes have access are radio/TV signals (but they do 
not have radios! in our case this would correspond, say, to neutrino-based communication 
technology), pollution/fire traces (almost not measurable and, if detected, explainable as natural 
fires) and the sights of artificial satellites and the international space station in the sky (which they 
would interpret also as a natural phenomena). Our civilization is almost not detectable by lost 
tribes! (This explains why they are lost!). Of course we can detect them, say by satellite search, 
but who will bother to contact them? Brazilian government estimates that there are two hundred 
not contacted populations inside Amazonia, and they will continue not contacted by a large time 
from now (again a persistence phenomenon).

	Persistence solves Fermi Paradox without appealing to non-testable sociological 
assumptions about galactic civilizations such as the Zoo hypothesis and the like. To explain the 
Great Silence we need not assume that we are inside a reserve preserved by the galactic ONU. It 
suffices that we are like an Indonesian tribe in a lost island (there are 17000 such islands but only 
6000 are inhabited, another example of the persistence phenomenon).  Persistence also explains 
the null results of SETI project when screening neighbor stars of the Sun. Indeed, it is a waste of 
time and money to examine neighboring stars, because if there were civilizations there then they 
should have colonized Earth a long time ago. SETI null results are fully compatible (indeed 
required) by the persistence phenomenon.

	To test these ideas we are presently performing simulations in the following model. We put 
$N$ random points in a two dimensional sheet. Each point has a binary state ($0 =$ unoccupied, $A = $
occupied by a technological civilization). Then, we define the following branching process:
\begin{eqnarray}
0 &\rightarrow&  A \:\:\: \mbox{with rate $\epsilon$  (emergence of civilizations)}\:,\nonumber\\
A &\rightarrow& 0 \:\:\:\mbox{with rate $q$  (decay of civilization)} \:,\nonumber\\
A &\rightarrow& A + A \:\:\:\mbox{with rate $p$ (colony formation)} \:, \nonumber\\
A &\rightarrow& A \:\:\:\mbox{with rate $r=1-q-p$} \:.
\end{eqnarray}

Since $\epsilon \ll p,q$, consider the case $\epsilon =0$ and the emergence of a non extensive
number of civilizations.
It is clear that, if the average branching ratio $\sigma = (1-q-p) + 2 p = 1+p-q$
is less than one, then the branching 
process is subcritical and the galaxy will not be colonized. 
This is the usual Fermi Paradox solution. 
However, even if  $\sigma > 1$, we must consider spatial correlations and density limiting
constraints of the colonization process (the exponential growth for  $sigma>1$ in
the above mean field calculation is only a first approximation).
Due to the persistence phenomenon, large regions will remain not 
colonized for considerable time intervals. A complete account of the simulation results of this 
model in d=2 and d=3 space 
(which is similar to a branching partial self-avoiding polymer model or a model of diffusion 
of ant nests) will be reported elsewhere \cite{KS}.

If the persistence solution to Fermi paradox is correct, then Earth belongs to a huge but 
poorly inhabited galactic domain. Since these regions could be very large (as the analogy with the 
highly non-uniform Earth global colonization suggests), the probability that Earth is inside them is 
high indeed. We must consider the distressing possibility that we live not in the 
``first world" part of 
the galaxy, but in a large region analogous to Amazonia, Indonesian islands or Africa. I conclude 
that Earth might not be a typical but an exotic place, being a persistent site unvisited by the galactic 
colonization process. We must all perform a change of view about what is typical and what is 
exotic (this fact motivated my early joke about Americans/Europeans and Brazilians). The 
opposite (Asimov/Tippler \cite{Crawford}) view that we are the first technological 
civilization of the galaxy or the 
``Star Trek" expectation that one day we will be the ``leaders" of the Federation of United Planets probably 
originates from a misplaced first world perspective, or better, it is simply wishful thinking: we 
``wish" that this could be true because we cannot accept the Copernican Mediocrity Principle that 
we pertain to a mediocre civilization. As a literary antidote to these naive views, I recommend the 
books of Stanislaw Lem \cite{Lem}.
 
{\bf Acknowledgments:} The author thanks Nelson Alves, Antonio Carlos Roque 
and Nestor Caticha for reading, commenting and pruning the manuscript, and Alexandre S. 
Martinez for calling attention to the nice Earth photography.

\end{document}